\documentclass[preprint,prx,amsmath,amssymb,longbibliography,superscriptaddress]{revtex4-1}

\pdfoutput=1

\usepackage{graphicx}
\usepackage{amsmath,latexsym,tabularx}
\usepackage{xcolor}
\usepackage{multirow}
\usepackage[
  colorlinks=true,
  urlcolor=blue,
  linkcolor=blue,
  citecolor=blue
]{hyperref}
\usepackage{multirow,booktabs}
\usepackage{float}
\usepackage{upgreek}
\usepackage{fancyhdr}

\usepackage{setspace}
\usepackage{caption}

\newcommand{\affilLL}[0]{MIT Lincoln Laboratory, Lexington, Massachusetts 02421, USA}

\newcommand{\specificthanks}[1]{\@fnsymbol{#1}}

\begin{document}

\title{\normalsize An Integrated Ultralow Noise Spiral Interferometric Laser}



\author{William Loh}
\thanks {Correspondence to william.loh@ll.mit.edu}

\affiliation{\affilLL}

\author{David Reens}
\affiliation{\affilLL}

\author{Dave Kharas}
\affiliation{\affilLL}

\author{Alkesh Sumant}
\affiliation{\affilLL}

\author{Connor Belanger}
\affiliation{\affilLL}

\author{Eli Briskin}
\affiliation{\affilLL}

\author{Dodd Gray}
\affiliation{\affilLL}

\author{Alexander Medeiros}
\affiliation{\affilLL}

\author{Ryan T. Maxson}
\affiliation{\affilLL}

\author{William Setzer}
\affiliation{\affilLL}

\author{Ethan Clements}
\affiliation{\affilLL}

\author{Wonseok Shin}
\affiliation{\affilLL}

\author{Paul W. Juodawlkis}
\affiliation{\affilLL}

\author{Cheryl Sorace-Agaskar}
\affiliation{\affilLL}

\author{Siva Yegnanarayanan}
\affiliation{\affilLL}

\author{Danielle Braje}
\affiliation{\affilLL}

\author{Robert McConnell}
\affiliation{\affilLL}

\begin{abstract}

Photonic integration offers the potential to bring complex high-performance optical systems to the form factor of a compact semiconductor chip. However, the range of system functions accessible critically depends on the extent to which free-space and fiber components can be made integrable. The ultralow-expansion cavity-stabilized laser---often used in precision metrology, high-resolution sensors, and advanced systems in atomic physics---is one component that currently has no direct parallel on chip. Lasers stabilized to photonically-integrated resonators exist, but exhibit considerably higher frequency noise and are accompanied by large levels of frequency drift. We demonstrate here a new architecture for an ultranarrow linewidth integrated laser based on stabilization to a sinusoidal fringe of an interferometer having a long 25-m unbalanced delay line. Our interferometric laser not only advances the state-of-the-art for on-chip lasers, but we in addition introduce an amplitude locking scheme that greatly suppresses the laser's long-term frequency wander. We achieve a record on-chip fractional frequency noise of $5.6 \times 10^{-14}$, corresponding to a linewidth of 12 Hz centered at 1348 nm. To showcase the utility of this laser, we divide the optical carrier to microwave frequencies, demonstrating the ability to outperform state-of-the-art quartz crystal oscillators by 15 dB or more.

\end{abstract}

\maketitle


Bulk reference cavities constructed from ultralow-expansion (ULE) glass \cite{Young1999, Notcutt2005, Ludlow2007, Leibrandt2011, Jiang2011} or cryogenic crystalline silicon \cite{Kessler2012, Robinson2019} form the basis for the most stable lasers in existence today. Such lasers are currently the backbone for numerous applications in basic and applied science, including those of trapped-ion quantum computers \cite{Haffner2008}, optical-atomic clocks based on a narrow-linewidth transition \cite{Hinkley2013, Bloom2014, Godun2014, Ludlow2015, Loh2020, Huntemann2016, Brewer2019}, ultralow-noise optical-to-microwave frequency synthesizers \cite{Fortier2011, Li2014, Kudelin2024, Sun2024, Loh2025}, Hertz-level precision spectrometers \cite{Rafac2000}, and high-resolution sensors \cite{Abbott2016, Marra2018}. As a condition for their extraordinary performance, these reference cavities require both active vacuum pumping and a considerable degree of temperature stabilization and isolation from the environment. The extensive stabilization combined with an already sizable optical cavity make cavity-stabilized lasers physically large and too unwieldy for use outside of the laboratory. Yet, in the context of future applications that demand scalability, portability, mass-manufacturability, and robustness, a laser capable of matching the performance of bulk cavity-stabilized lasers while maintaining integrability with other system components on chip remains a highly desirable goal.

Over the last decade, substantial effort has been devoted to developing compact low-noise lasers based upon either miniaturizing the configuration of existing ULE cavities \cite{Davila-Rodriguez2017, Didier2019, Jin2022, Liu2024} or directly establishing optical cavities on chip for laser stabilization \cite{Lee2012, Loh2015, Gundavarapu2019, Xiang2023}. On-chip cavities, while attractive due to their amenability for system integration, face an increased set of challenges due to (1) thermal noise and drift in the waveguiding material, (2) lower quality factors (Q) resulting from higher propagation loss, and (3) large levels of intensity-induced frequency noise. To date, integrated lasers have showcased exceptionally low levels of intrinsic/fundamental linewidth derived from offset frequencies far removed from the carrier. Yet for lower offset frequencies most relevant to real-world applications, the noise often increases sharply, which considerably degrades the laser's actual linewidth. The current record performance across all integrated laser technologies was achieved by a 6.1-meter spiral cavity, demonstrating a fractional frequency noise of $7.5 \times 10^{-14}$ and a linewidth of 16.7 Hz at 1348 nm wavelength \cite{WLoh2025}. The long spiral waveguide \cite{Lee2013, Guo2022, Liu2022, He2024} effectively averages down the laser's thermorefractive noise, while the ultralow optical losses greatly narrow the width of the cavity's resonances. However, as a consequence of this optimized geometry, the spiral cavity architecture also reaches a limit where further improvements are possible only if the physical spiral length achievable on chip, the quality factor of the resonances, and the frequency drift at longer time scales can be improved simultaneously.

We show here a fundamentally different approach to on-chip laser stabilization, based on using an unbalanced Mach-Zehnder (MZ) interferometer in lieu of a traditional resonant cavity to serve as the master reference. The structure of our interferometer consists of two integrated waveguide couplers with 25 meters of excess spiral waveguide delay on one interferometer arm. Such delay-line interferometers have also been previously implemented in fiber platforms \cite{Kefelian2009, Jeon2025} and have demonstrated levels of performance rivaling that of cavity-stabilized lasers, but with the caveat of requiring 1 km or more of fiber delay length. A key question remains as to whether chip-integrated interferometers, with their substantially shorter delay lengths, would be viable for laser stabilization. Our work here not only affirms this to be true, but demonstrates that this scheme surpasses the best on-chip stabilization methods of today, thereby enabling continued future scaling of integrated lasers to higher levels of performance. In comparison to prior endeavors in utilizing on-chip interferometers for laser stabilization \cite{Idjadi2024, Cheng2025}, our approach solves two important limitations. First, we avoid the use of any resonant structures within the interferometer (ring resonators, spiral resonators, etc), whose added noise would inadvertently limit the interferometer stability to that of the cavity. Second, owing to our ultralow optical losses, we are able to maximize the spiral mode volume that can be laid out on chip and achieve delay lengths of 25 m, $2800 \times$ larger than that previously reported. Beyond these improvements, we introduce an amplitude locking scheme specific to the interferometer architecture that stabilizes its long-term frequency drift by an order of magnitude, down to the level of 24 Hz/s. Altogether, we set a new record for the stability achievable by an integrated laser of $5.6 \times 10^{-14}$, which we then use to perform optical frequency division down to 10 GHz and showcase phase noise 15 dB or more better than state-of-the-art quartz crystal oscillators.

\section{Results}


\begin{figure}[t b !]
\includegraphics[width = 0.95 \columnwidth]{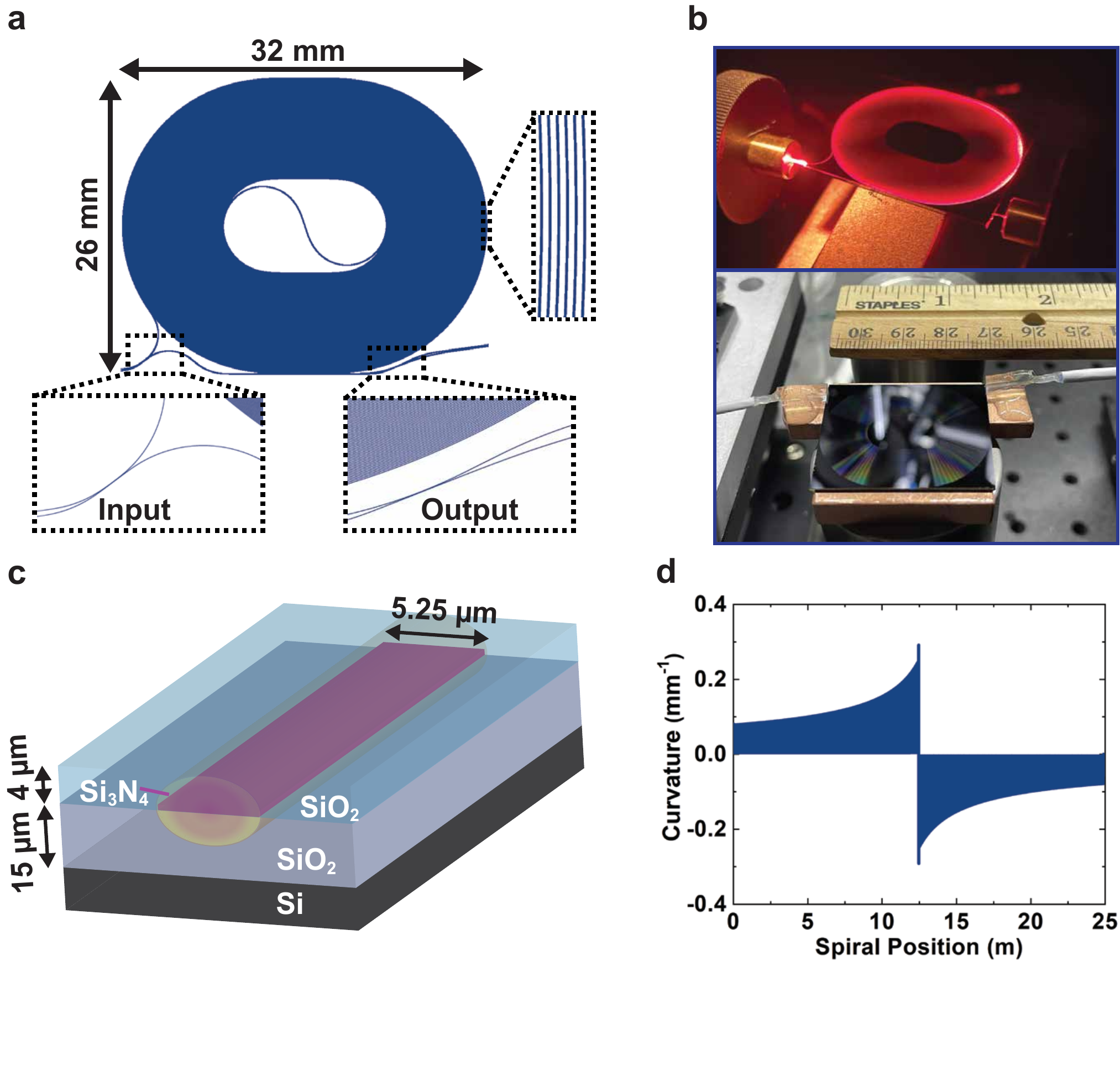}
\caption{
    \textbf{Spiral MZ Interferometer Configuration}
    \textbf{a}, Layout of a 25-m spiral MZ interferometer on a 2.6 cm $\times$ 2.6 cm semiconductor chip. The zoomed-in sections show the input ports (lower left), output ports (lower right), and the individual waveguide traces of the interferometer.
    \textbf{b}, Photographs of the spiral MZ interferometer. The top photograph shows the scatter out of the spiral delay line when red light sent is through the input port. The bottom photograph depicts the spiral MZ interferometer with fiber arrays bonded to the input and output ports.
    \textbf{c}, Spiral waveguide cross section depicting a Si$_3$N$_4$ core region bounded above and below by SiO$_2$ cladding. The optical mode is illustratively shown to be loosely confined by the Si$_3$N$_4$, with most of the power guided in the SiO$_2$.
    \textbf{d}, Plot of the curvature along the path of the spiral delay line. Over the 25-m length of the spiral, the minimum bending radius is 3.4 mm.
}
\label{fig:fig1}
\end{figure}

An interferometer offers many important advantages in comparison to a resonator for laser stabilization---the foremost being that the configuration of an interferometer can readily make use of balanced detection. Figure 1a depicts a schematic of our spiral MZ interferometer, which fits on a 2.6 cm $\times$ 3.2 cm chip, and includes two input ports on the left and two output ports on the right (see Methods A for more details). From input to output, the interferometer transfer function forms a sinusoid whose slope is proportional to the length of the unbalanced delay. Sending the outputs to a pair of balanced detectors not only achieves twice the interferometer slope, but more importantly, enables a large degree of cancellation of the interrogating laser's common-mode relative intensity noise (RIN). The interferometer slope converts frequency fluctuations of the interrogating laser to a detectable amplitude. This amplitude compared to the level of background noise determines the degree to which the laser lock can accurately track the sinusoidal fringe. With the use of balanced detection, the signal-to-noise (SNR) increases considerably, which enables the interferometer to markedly surpass the performance of otherwise higher-Q optical cavities.

The interferometer architecture has a few other notable advantages worth mentioning. First, because the locking is directly accomplished using a fringe of the sinusoidal interference pattern, no phase modulator is needed---simplifying the locking circuitry and circumventing the dominant source of residual amplitude modulation induced frequency drift \cite{Zhang2014}. Second, since the interference occurs at the output coupler, the optical power is constant for the vast majority of the interferometer structure. This largely prevents frequency fluctuations from unintentionally converting to amplitude fluctuations, which would otherwise reconvert as additional frequency fluctuations at a later stage. Third, because the optical power measured at the output is directly correlated to the power that traverses through the long interferometer delay line, we are able to implement straightforward amplitude locking schemes to considerably stabilize the interferometer's long-term drift. Such techniques are not easily implemented for ring or spiral resonators since their output becomes a complex superposition of both the field that bypasses the cavity and the intracavity field that couples out. Fourth, the spiral structure is better suited to single-pass waveguides compared to resonant cavities, due to the tight bends required to complete the cavity loop \cite{WLoh2025}. This enables us to achieve a spiral delay of 25-m on chip despite utilizing a diffuse optical mode. Lastly, the interferometer architecture does not suffer from loss in the same capacity as optical cavities, where increases in loss would sharply degrade the cavity Q. In an interferometer, loss decreases both the signal and the level of noise at the output, causing the SNR to be maintained provided the noise is above the shot noise limit. 

Figure 1b shows two photographs of the fabricated spiral MZ interferometer. The top photograph depicts the spiral when red light is sent through the top input port. Despite operating at half the intended wavelength, the red light makes its way across much of the spiral as indicated by the visible light scatter. The bottom photograph shows the spiral with fiber arrays bonded to the input and output waveguides. Each of the fiber arrays consists of two fibers spaced by 127 $\mu$m and angled at 8$^{\circ}$. Though the fiber arrays allow for light to be sent along two possible input ports, we primarily use the intended (top) port to maximize the depth of the sinusoidal interference fringe.

Our spiral waveguide cross section (Fig. 1c) consists of 40 nm of Si$_3$N$_4$ as the core layer having a width of 5.25 $\mu$m , along with 4 $\mu$m and 15 $\mu$m of SiO$_2$ above and below to serve as the cladding, respectively. From simulations, we determine this structure to enable single-mode operation at 1348 nm with a large diffuse 13.4 $\mu$m$^2$ optical mode. This waveguide cross section is maintained along the entire spiral length of 25 meters, with curvature gradually increasing towards the center of the spiral (Fig. 1d). Near the center, the radius of curvature starts at 3.4 mm for the sinusoidal bend, and then increases to 4 mm and larger for the remaining laps of the spiral. This configuration enables low optical losses of 0.15 dB/m averaged across the length of the spiral.


\begin{figure}[t b !]
\includegraphics[width = 0.95 \columnwidth]{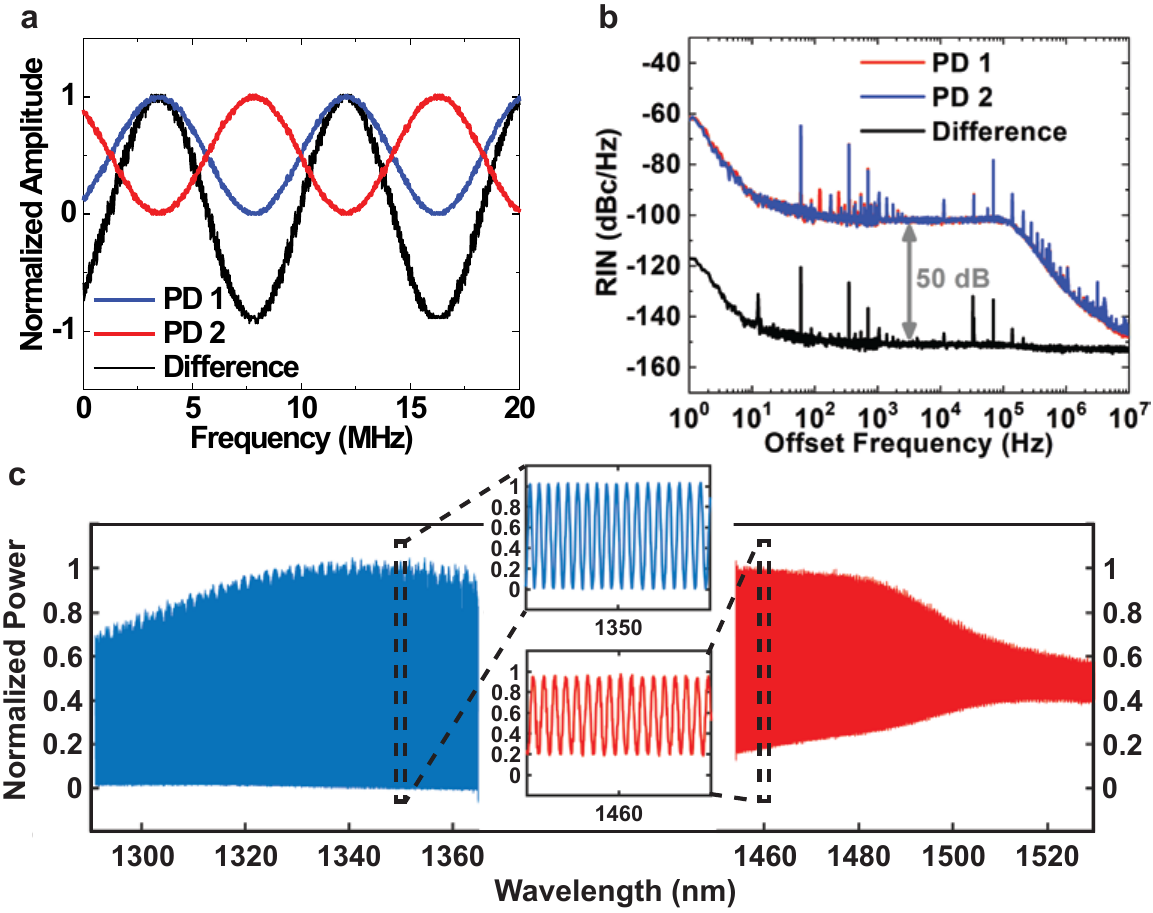}
\caption{
    \textbf{Spiral MZ Interferometer Characterization}
    \textbf{a}, Sinusoidal interference pattern measured by two independent photodetectors (PD) tracking the outputs of the spiral MZ interferometer. Balanced detection doubles the slope of the interference fringe.
    \textbf{b}, Measured RIN of the interrogating laser prior to laser stabilization. Balanced photodetection suppresses the measured RIN by 50 dB.
    \textbf{c}, Broadband scan of the interferometer response from 1290 nm -- 1365 nm and 1454 nm -- 1530 nm. The zoomed-in sections near 1350 nm and 1460 nm show the sinusoidal interference pattern to be preserved across a broad wavelength range.
}
\label{fig:fig2}
\end{figure}

From the input port to either of its two output ports, the spiral MZ interferometer traces out a sinusoidal response whose period varies inversely with the length of the spiral delay. Figure 2a shows the interference pattern measured at each of the photodetected outputs with the peak amplitudes normalized to unity. For 25 meters of spiral delay length, the interference period is 8 MHz. Furthermore, the measured extinction is 19.5 dB on each of the two output ports, indicating that the fields traversing each of the interferometer arms are well matched in amplitude. Note that the first splitter is intentionally offset from an even split to compensate for the loss of the 25-meter spiral. After subtracting the outputs in a balanced photodetection scheme, the sinusoidal interference pattern doubles in amplitude while maintaining the same periodicity. The interferometer response has an effective linewidth of 4 MHz corresponding to a Q of 55.6 million, which is a consequence of the sinusoid's 50$\%$ duty cycle (see Methods B). This may appear low compared to the $>$ 100 million Qs already achievable by state-of-the-art integrated resonators today. However, the SNR of our interferometer locking scheme here exceeds by orders of magnitude the highest performing Pound-Drever-Hall stabilization efforts of integrated resonators, due to the suppression of noise via balanced detection.

Figure 2b showcases the measured RIN of the interrogating laser with and without balanced detection. The interrogating laser includes a semiconductor optical amplifier (SOA) that provides independent amplitude control of the laser in our locking scheme. Each of the individual photodetectors measures the same level of RIN, which becomes suppressed by 50 dB when their difference is measured via balanced detection. This suppressed RIN, in conjunction with the slope of the interferometer's sinsoidal response, directly determine the frequency noise performance of the stabilized laser (see Methods C).

The broadband operation of the spiral MZ interferometer is illustrated in Fig. 2c. The interferometer is probed with a tunable laser whose wavelength is swept from 1290 nm to 1365 nm; we measure the signal from one of the two output ports chosen arbitrarily. A zoom-in around the 1350 nm region shows the sinusoidal pattern to exhibit nearly full extinction of the interferometer response. This extinction is observed to weaken only slightly at the shorter wavelength side, suggesting that the spiral MZ interferometer may be used effectively for laser stabilization across the entirety of the wavelength range tested. In addition, a second tunable laser covering the range of 1454 nm to 1530 nm was used to probe the same ports of the spiral MZ interferometer. This range of wavelengths falls well outside the intended operating wavelength of 1348 nm, but still provides useful information on the interferometer's broadband operation. The zoom-in around 1460 nm indicates that the extinction has decreased to $\sim80 \%$ at a wavelength over 110 nm away from the intended operating point. From the measured response, the spiral MZ interferometer appears to yield a suitable fringe for locking up to $\sim$1500 nm in wavelength.


\begin{figure}[t b !]
\includegraphics[width = 0.95 \columnwidth]{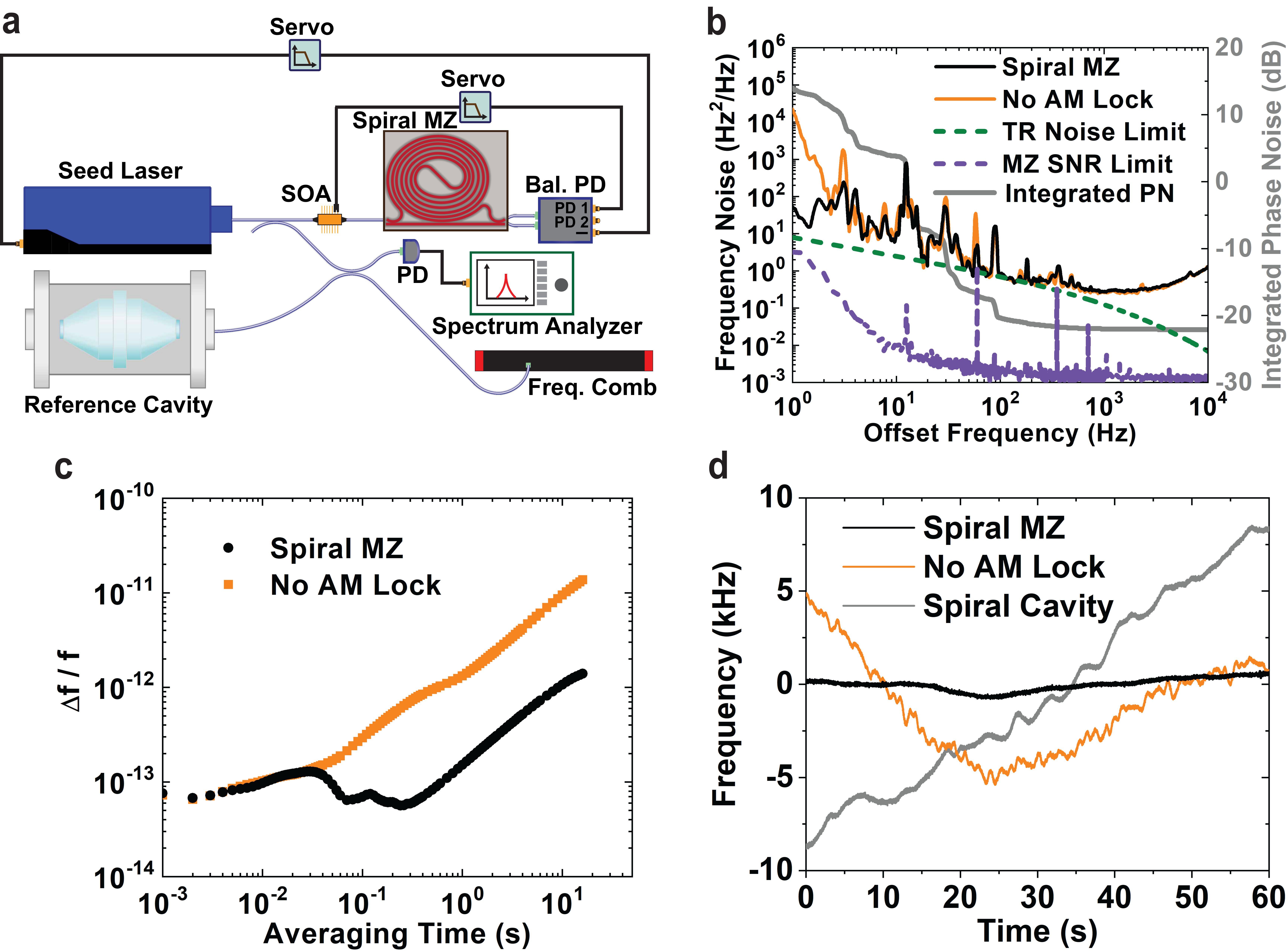}
\caption{
    \textbf{Laser Stabilization to Spiral MZ Interferometer}
    \textbf{a}, Schematic of a spiral interferometric laser consisting of an amplified seed laser that probes the interferometer chip. Various components support the locking of the seed laser to the interferometer and also the subsequent measurement of the stabilized laser performance. PD, photodetector; Bal. PD, balanced photodetector; Freq. Comb, frequency comb.
    \textbf{b}, Measured frequency noise of the spiral interferometric laser. Also shown for comparison are the laser operating without the amplitude modulation (AM) lock, the thermorefractive (TR) noise limit for a 25-meter spiral, the performance limit derived from the laser SNR, and the value of the integrated phase noise (PN) from higher to lower offset frequencies.
    \textbf{c}, Fractional frequency noise of the spiral interferometric laser with and without the use of AM locking. The AM lock primarily affects the laser's long-term frequency drift.
    \textbf{d}, Measurement of the laser frequency for three different laser configurations across a one minute time span. The spiral interferometric laser is compared with and without the use of the AM lock. The current best-performing chip-integrated laser is also graphed, whose operation utilizes a 6.1-m spiral cavity for stabilization.
}
\label{fig:fig3}
\end{figure}

Our spiral interferometric laser system consists of a 1348 nm seed laser that is amplified by an SOA and routed to the interferometer's top input port (Fig. 3a). The light splits in the interferometer chip and subsequently recombines, and the two interferometer output ports are then detected off-chip on a balanced photodetector. The difference photocurrent is used to lock the seed laser to the sinusoidal fringe of the interferometer. The balanced photodetector also provides access to the individual outputs of each photodetector, which we utilize to feed back on the SOA amplitude to stabilize the optical power that traverses the spiral interferometer. A splitter after the seed laser divides the power in two for comparing with a ULE reference cavity. The combined outputs are sent to a spectrum analyzer for noise measurement and separately to a self-referenced frequency comb for division to microwave frequencies. We note that our implementation of the amplitude lock does not present any conflicts with the laser lock to the interferometer fringe, whose error signal is also based on amplitude. Since the laser lock is operated at a zero crossing, any non-zero difference photocurrent measured prompts the seed laser to change its frequency until the error reaches zero. This occurs at a unique point as an amplitude change on the SOA is unable to bring the difference photocurrent to zero. Once the seed laser frequency is locked, the amplitude lock then acts to stabilize the waveguide power. 

The measured frequency noise of the spiral interferometric laser (Fig. 3b) indicates that much of the spectrum is at or near the thermorefractive noise limit \cite{Gorodetsky2004, Matsko2007, Huang2019, Panuski2020} derived for a 25-m spiral waveguide. At the higher offset frequencies, the servo bandwidth of $\sim$500 kHz results in a slight increase in noise as the feedback gain diminishes. At lower offset frequencies, the noise is slightly above the thermorefractive limit, constrained primarily by system vibration. These measured levels are 6 dB lower across the majority of the spectrum compared to the best frequency noise demonstrated by on-chip optical cavities \cite{WLoh2025}, owing to a reduced thermorefractive noise and a greatly increased SNR of the laser lock. The SNR limit is plotted in Fig. 3b, calculated from the measured RIN and the sinusoidal conversion slope of the interferometer. This limit indicates that close to an additional 30 dB improvement in the noise performance is supported by the interferometric lock provided that the thermorefractive noise could be further reduced. Without the amplitude lock engaged, the spiral interferometric laser's frequency noise increases sharply at low offset frequencies below 3 Hz (see Methods D for operational details when the lock is detuned from the center). By integrating the laser's phase noise, we estimate the laser's linewidth to be 12 Hz.

The fractional frequency noise over time of the spiral interferometric laser (Fig. 3c) provides more detailed information on the laser's performance at longer time scales. The measured fractional frequency noise starts at $7.6 \times 10^{-14}$ at 1 ms and reaches $5.6 \times 10^{-14}$ by 240 ms. These measured values also considerably improve over the best demonstrations of integrated optical resonators, which have achieved $1.6 \times 10^{-13}$ at 1 ms and a minimum of $7.5 \times 10^{-14}$ at 30 ms. From the fractional frequency noise, we extract a linewidth of 12 Hz for the spiral interferometric laser, which compares well with the value of the linewidth derived earlier from the integrated phase noise. The amplitude lock reduces the long-term drift to the level of 24 Hz/s, $>10\times$ better than previously demonstrated by on-chip resonators. When the amplitude lock is not engaged, the fractional frequency noise at longer time scales ($>30$ ms) degrades by a factor of 9.

The spiral interferometric laser's frequency stability is better visualized when its frequency is plotted as a function of time, sampled at 1 ms intervals (Fig. 3d). Over the course of one minute, the spiral interferometric laser makes a maximum excursion of $\pm600$ Hz from its center (see Methods E and F for details of the MZ interferometer's thermal response). In contrast, not only is the frequency drift noticeably larger without the use of the amplitude lock, but the faster fluctuations in frequency are more prominent. The prior best performing on-chip laser, which uses a spiral resonator, is also plotted for reference \cite{WLoh2025}. The frequency fluctuations are similarly much larger in comparison to the locked spiral interferometric laser, and the laser drifts 17 kHz over the same one minute time period.


\begin{figure}[t b !]
\includegraphics[width = 0.95 \columnwidth]{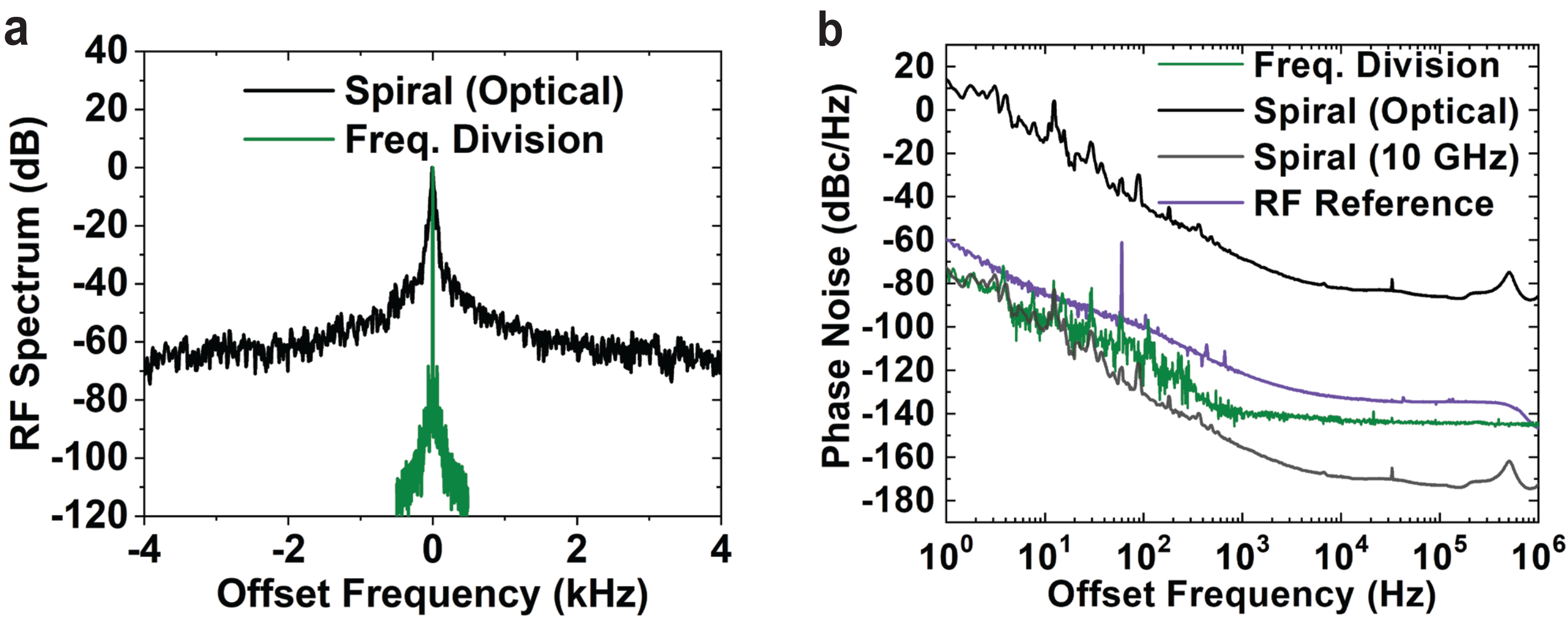}
\caption{
    \textbf{Optical Frequency Division of Spiral Interferometric Laser}
    \textbf{a}, Spectrum of the laser lineshape before and after optical frequency division to 10 GHz. The resolution bandwidth is 5 Hz and 1 Hz for the optical spectrum and the divided-down microwave spectrum, respectively.
    \textbf{b}, Measured phase noise of the spiral interferometric laser after optical frequency division to 10 GHz. Also shown are the measured phase noise at optical frequencies and projected to 10 GHz, as well as the phase noise of an ultralow noise oven-controlled crystal oscillator reference (Rohde $\&$ Schwarz SMB100B-B711).
}
\label{fig:fig4}
\end{figure}

We transfer the ultralow frequency noise achieved by the spiral interferometric laser to microwave frequencies by coherently dividing down the optical carrier. For division from 222 THz (1348 nm) to 10 GHz, the phase noise spectral density reduces by 87 dB. The frequency division is performed using an Er-fiber optical frequency comb (Vescent Photonics RUBRIComb) that is both self-referenced and has its comb teeth locked to the output of the spiral interferometric laser. The stabilized comb light is photodetected on a high-speed detector and bandpass filtered to provide an ultralow-noise microwave output at 10 GHz. Figure 4a shows the measured divided-down spectrum compared to the spectrum of the optical beat against a ULE reference cavity. The optical spectrum indicates a full-width half-maximum of 19 Hz, corresponding well with linewidths extracted from measurements of the laser noise after accounting for the 0.8-second sweep time. The divided-down 10 GHz microwave spectrum showcases the orders-of-magnitude reduction in noise achievable by optical frequency division. However, the linewidth after division is not quantifiable due to the spectrum's resolution bandwidth limit of 1 Hz. 

Figure 4b shows the measured phase noise of the spiral interferometric laser before and after optical frequency division. The phase noise at optical frequencies provides a reference point for the expected performance at 10 GHz when ideal division is performed. A servo bump is observed at $\sim500$ kHz offset frequency, which transfers over to the projected 10 GHz microwave phase noise, but at a level that is 87 dB lower. While the phase noise at 10 GHz is predicted to be below -170 dBc/Hz for offset frequencies above 10 kHz, additional sources of noise such as shot noise and thermal noise will limit the value of the floor achievable. Comparing to the measured 10 GHz signal after optical frequency division, we find that the divided-down microwave follows the projected performance at lower offset frequencies below 10 Hz. Above 10 Hz however, the measured phase noise diverges from projections, due primarily to reaching the measurement limit of the phase noise analyzer for offset frequencies 10 Hz -- 1 kHz and the thermal noise limit for offset frequencies $>1$ kHz. With a higher power photodetector, the thermal limit may be substantially improved upon. On the other hand, the phase noise measurement was performed with 1000-sample cross-correlation averaging across a 3-hour time period, and its limit was confirmed by sending light from a divided-down ULE reference cavity into the measurement system. Despite the extensive cross-correlation averaging, the level of phase-noise after dividing down the spiral interferometric laser was too low to fully resolve for much of the spectrum. For the measurable offset frequencies of 1 Hz and 10 Hz, we report at 10 GHz phase noise levels of -76 dBc/Hz and -99 dBc/Hz, respectively. These values are the lowest to date achieved from frequency division of a chip-integrated optical reference. Compared to a state-of-the-art ultralow phase noise oven-controlled crystal oscillator reference (Fig. 4b), we find our achieved phase noise levels to be improved by 15 dB or more across the spectrum.

\section{Discussion}

Our spiral interferometric laser represents an important step in enabling integrated lasers to continue scaling their performance to narrower linewidths and lower noise. This demonstration comes at a time when traditional on-chip cavities have reached a limit in terms of both the cavity Q and the amount of waveguide length that can be spiraled onto a single reticle. Using our developed amplitude lock, the spiral interferometric laser considerably outperforms the best reports of integrated cavities to date, improving both the shorter-term frequency noise by 6 dB and the longer-term drift by an order of magnitude. Furthermore, upon optical frequency division of the spiral interferometric laser to 10 GHz, we showcase the lowest phase noise of a microwave signal derived from an integrated optical reference. Beyond future improvements to the interferometer architecture and the eventual integration of the seed laser and photodetectors on chip, the spiral interferometric laser may see immediate use in advancing the current capabilities of portable optical atomic clocks and high precision sensors and spectrometers. Once coherently divided down, the synthesized microwave outperforms conventional RF oscillators and may lead to future capabilities in communications networks and radar exceeding what is possible today. Finally, with the interferometer's potential for broadband operation and its reduced dependence on waveguide loss, we anticipate that ultranarrow-linewidth visible wavelength lasers will soon become readily accessible.

\section{Acknowledgements}

We acknowledge assistance from Henry Timmers, Cole Smith, and Nate Phillips with the frequency comb setup. We acknowledge John Chiaverini for assistance with the manuscript preparation. This material is based upon work supported by the  Under Secretary of Defense for Research and Engineering under Air Force Contract No. FA8702-15-D-0001 or FA8702-25-D-B002. Any opinions, findings, conclusions or recommendations expressed in this material are those of the author(s) and do not necessarily reflect the views of the  Under Secretary of Defense for Research and Engineering. DISTRIBUTION STATEMENT A. Approved for public release. Distribution is unlimited.

\section{Contributions}

W.L., D.R., and R. M. conceived, designed and carried out the experiments with the spiral interferometric laser.  W.L. and S. Y. conceived, designed and carried out the experiments with the optical frequency division. D.K. and A.S. fabricated the spiral resonators. C.B., W. L., A. M., and R.T.M. performed the integrated chip fiber attach. E. B., D. G., and W. L. performed the broadband measurements of the MZ interferometer. All authors discussed the results and contributed to the manuscript.

\section{Competing interests}

The authors declare no competing interests.

\section{References}
\bibliography{Spiral_MZ_biblio}

@article{Hinkley2013,
  title = {An atomic clock with $10^{-18}$ instability},
  author = {Hinkley, N. and Sherman, J. A. and Philips, N. B. and Schioppo, M. and Lemke, N. D. and Beloy, K. and Pizzocaro, M. and Oates, C. W. and Ludlow, A. D.},
  journal = {Science},
  volume = {341},
  issue = {6151},
  year = {2013},
  pages = {1215-1218},
  doi = {10.1126/science.1240420},
  url = {https://doi.org/10.1126/science.1240420}
}

@article{Bloom2014,
  title = {An optical lattice clock with accuracy and stability at the $10^{-18}$ level},
  author = {Bloom, B. J. and Nicholson, T. L. and Williams, J. R. and Campbell, S. L. and Bishof, M. and Zhang, X. and Bromley, S. L. and Ye, J.},
  journal = {Nature},
  volume = {506},
  year = {2014},
  pages = {71-75},
  publisher = {Nature Research},
  doi = {10.1038/nature12941},
  url = {https://doi.org/10.1038/nature12941}
}

@article{Brewer2019,
  title = {$^{27}\mathrm{Al}^{+}$ Quantum-logic clock with a systematic uncertainty below $10^{-18}$},
  author = {Brewer, S. M. and Chen, J. -S. and Hankin, A. M. and Clements, E. R. and Chou, C. W. and Wineland, D. J. and Hume, D. B. and Leibrandt, D. R.},
  journal = {Phys. Rev. Lett.},
  volume = {123},
  issue = {3},
  year = {2019},
  pages = {033201},
  doi = {10.1103/PhysRevLett.123.033201},
  url = {https://doi.org/10.1103/PhysRevLett.123.033201}
}

@article{Ludlow2015,
  title = {Optical atomic clocks},
  author = {Ludlow, A. D. and Boyd, M. M. and Ye, J. and Peik, E. and Schmidt, P. O.},
  journal = {Rev. Mod. Phys.},
  volume = {87},
  issue = {2},
  year = {2015},
  pages = {637},
  doi = {10.1103/RevModPhys.87.637},
  url = {https://doi.org/10.1103/RevModPhys.87.637}
}

@article{Godun2014,
  title = {Frequency ratio of two optical clock transitions in $^{171}\mathrm{Yb}^{+}$ and constraints on the time variation of fundamental constants},
  author = {Godun, R. M. and Nisbet-Jones, P. B. R. and Jones, J. M. and King, S. A. and Johnson, L. A. M. and Margolis, H. S. and Szymaniec, K. and Lea, S. N. and Bongs, K. and Gill, P.},
  journal = {Phys. Rev. Lett.},
  volume = {113},
  issue = {21},
  year = {2014},
  pages = {210801},
  doi = {10.1103/PhysRevLett.113.210801},
  url = {https://doi.org/10.1103/PhysRevLett.113.210801}
}

@article{Huntemann2016,
  title = {Single-ion atomic clock with $3\times10^{-18}$ systematic uncertainty},
  author = {Huntemann, N. and Sanner, C. and Lipphardt, B. and Tamm, Chr. and Peik, E.},
  journal = {Phys. Rev. Lett.},
  volume = {116},
  issue = {6},
  year = {2016},
  pages = {063001},
  doi = {10.1103/PhysRevLett.116.063001},
  url = {https://doi.org/10.1103/PhysRevLett.116.063001}
}

@article{Abbott2016,
  title = {Observation of Gravitational Waves from a Binary Black Hole Merger},
  author = {Abbott, B. P. and Abbott, R. and Abbott, T. D. and Abernathy, M. R. and Acernese, F. and Ackley, K. and Adams, C. and Adams, T. and Addesso, P. and Adhikari, R. X. and Adya, V. B. and Affeldt, C. and Agathos, M. and Agatsuma, K. and Aggarwal, N. and others},
  collaboration = {{LIGO} Scientific Collaboration and Virgo Collaboration},
  journal = {Phys. Rev. Lett.},
  volume = {116},
  issue = {6},
  pages = {061102},
  numpages = {16},
  year = {2016},
  month = {Feb},
  publisher = {American Physical Society},
  doi = {10.1103/PhysRevLett.116.061102},
  url = {https://link.aps.org/doi/10.1103/PhysRevLett.116.061102}
}

@article{Young1999,
  title = {Visible Lasers with Subhertz Linewidths},
  author = {Young, B. C. and Cruz, F. C. and Itano, W. M. and Bergquist, J. C.},
  journal = {Phys. Rev. Lett.},
  volume = {82},
  issue = {19},
  pages = {3799--3802},
  numpages = {0},
  year = {1999},
  month = {May},
  publisher = {American Physical Society},
  doi = {10.1103/PhysRevLett.82.3799},
  url = {https://link.aps.org/doi/10.1103/PhysRevLett.82.3799}
}

@article{Jiang2011,
  title = {Making optical atomic clocks more stable with $10^{-16}$-level laser stabilization},
  author = {Jiang, Y. Y. and Ludlow, A. D. and Lemke, A. D. and Fox, R. W. and Sherman, J. A. and Ma, L. -S. and Oates, C. W.},
  journal = {Nat. Photon.},
  volume = {5},
  year = {2011},
  pages = {158-161},
  publisher = {Nature Research},
  doi = {10.1038/nphoton.2010.313},
  url = {https://doi.org/10.1038/nphoton.2010.313}
}

@article{Kessler2012,
  title = {A sub-40-{mHz}-linewidth laser based on a silicon single-crystal optical cavity},
  author = {Kessler, T. and Hagemann, C. and Grebing, C. and Legero, T. and Sterr, U. and Riehle, F. and Martin, M. J. and Chen, L. and Ye, J.},
  journal = {Nat. Photon.},
  volume = {6},
  year = {2012},
  pages = {687-692},
  publisher = {Nature Research},
  doi = {10.1038/nphoton.2012.217},
  url = {https://doi.org/10.1038/nphoton.2012.217}
}

@article{Lee2012,
  title = {Chemically etched ultrahigh-{Q} wedge-resonator on a silicon chip},
  author = {Lee, H. and Chen, T. and Li, J. and Yang, K. Y. and Jeon, S. and Painter, O. and Vahala, K. J.},
  journal = {Nat. Photon.},
  volume = {6},
  year = {2012},
  pages = {369-373},
  publisher = {Nature Research},
  doi = {10.1038/nphoton.2012.109},
  url = {https://doi.org/10.1038/nphoton.2012.109}
}

@article{Loh2015,
  title = {Dual-microcavity narrow-linewidth Brillouin laser},
  author = {Loh, W. and Green, A. A. S. and Baynes, F. N. and Cole, D. C. and Quinlan, F. J. and Lee, H. and Vahala, K. J. and Papp, S. B. and Diddams, S. A.},
  journal = {Optica},
  number = {3},
  volume = {2},
  month = {Mar},
  year = {2015},
  pages = {225--232},
  publisher = {OSA},
  doi = {10.1364/OPTICA.2.000225},
  url = {https://doi.org/10.1364/OPTICA.2.000225}
}

@article{Gundavarapu2019,
  title = {Sub-hertz fundamental linewidth photonic integrated Brillouin laser},
  author = {Gundavarapu, S. and Brodnik, G. M. and Puckett, M. and Huffman, T. and Bose, D. and Behunin, R. and Wu, J. and Qiu, T. and Pinho, C. and Chauhan, N. and Nohava, J. and Rakich, P. T. and Nelson, K. D. and Salit, M. and Blumenthal, D. J.},
  journal = {Nat. Photon.},
  volume = {13},
  year = {2019},
  pages = {60-67},
  publisher = {Nature Research},
  doi = {10.1038/s41566-018-0313-2},
  url = {https://doi.org/10.1038/s41566-018-0313-2}
}

@article{Leibrandt2011,
author = {Leibrandt, D. R. and Thorpe, M. J. and Bergquist, J. C. and Rosenband, T.},
journal = {Opt. Express},
number = {11},
pages = {10278--10286},
publisher = {OSA},
title = {Field-test of a robust, portable, frequency-stable laser},
volume = {19},
month = {May},
year = {2011},
url = {http://www.opticsexpress.org/abstract.cfm?URI=oe-19-11-10278},
doi = {10.1364/OE.19.010278}
}

@article{Davila-Rodriguez2017,
author = {Davila-Rodriguez, J. and Baynes, F. N. and Ludlow, A. D. and Fortier, T. M. and Leopardi, H. and Diddams, S. A. and Quinlan, F.},
journal = {Opt. Lett.},
number = {7},
pages = {1277--1280},
publisher = {OSA},
title = {Compact, thermal-noise-limited reference cavity for ultra-low-noise microwave generation},
volume = {42},
month = {Apr},
year = {2017},
doi = {10.1364/OL.42.001277},
url = {http://ol.osa.org/abstract.cfm?URI=ol-42-7-1277}
}

@article{Didier2019,
  title = {Ultracompact reference ultralow expansion glass cavity},
  author={Didier, A. and Millo, J. and Marechal, B. and Rocher, C. and Rubiola, E. and Lecomte, R. and Ouisse, M. and Delporte, J. and Lacro\^{u}te, C. and Kersal\'{e}, Y.},
  journal = {Appl. Opt.},
  number = {22},
  pages = {6470--6473},
  publisher = {OSA},
  volume = {57},
  month = {Aug},
  year = {2018},
  url = {http://ao.osa.org/abstract.cfm?URI=ao-57-22-6470},
  doi = {10.1364/AO.57.006470},
}

@article{Robinson2019,
author = {J. M. Robinson and E. Oelker and W. R. Milner and W. Zhang and T. Legero and D. G. Matei and F. Riehle and U. Sterr and J. Ye},
journal = {Optica},
number = {2},
pages = {240--243},
publisher = {Optica Publishing Group},
title = {Crystalline optical cavity at 4{K} with thermal-noise-limited instability and ultralow drift},
volume = {6},
month = {Feb},
year = {2019},
url = {https://opg.optica.org/optica/abstract.cfm?URI=optica-6-2-240},
doi = {10.1364/OPTICA.6.000240},
}

@article{Loh2020,
author={W. Loh and J. Stuart and D. Reens and C. D. Bruzewicz and D. Braje and J. Chiaverini and P. W. Juodawlkis and J. M. Sage and R. McConnell},
journal={Nature},
year={2020},
volume={588},
pages={244--249},
title={Operation of an optical atomic clock with a Brillouin laser subsystem},
doi={10.1038/s41586-020-2981-6}
}

@article{Gorodetsky2004,
author = {M. L. Gorodetsky and I. S. Grudinin},
journal = {J. Opt. Soc. Am. B},
number = {4},
pages = {697--705},
publisher = {OSA},
title = {Fundamental thermal fluctuations in microspheres},
volume = {21},
month = {Apr},
year = {2004},
url = {http://josab.osa.org/abstract.cfm?URI=josab-21-4-697},
doi = {10.1364/JOSAB.21.000697},
}

@article{Matsko2007,
author = {Andrey B. Matsko and Anatoliy A. Savchenkov and Nan Yu and Lute Maleki},
journal = {J. Opt. Soc. Am. B},
number = {6},
pages = {1324--1335},
publisher = {OSA},
title = {Whispering-gallery-mode resonators as frequency references. {I}. Fundamental limitations},
volume = {24},
month = {Jun},
year = {2007},
url = {http://josab.osa.org/abstract.cfm?URI=josab-24-6-1324},
doi = {10.1364/JOSAB.24.001324},
}

@article{Huang2019,
title = {Thermorefractive noise in silicon-nitride microresonators},
author = {G. Huang and E. Lucas and J. Liu and A. S. Raja and G. Lihachev and M. L. Gorodetsky and N. J. Engelsen and T. J. Kippenberg},
journal = {Phys. Rev. A},
volume = {99},
issue = {6},
pages = {061801},
numpages = {5},
year = {2019},
month = {Jun},
publisher = {American Physical Society},
doi = {10.1103/PhysRevA.99.061801},
url = {https://link.aps.org/doi/10.1103/PhysRevA.99.061801}
}

@article{Panuski2020,
title = {Fundamental Thermal Noise Limits for Optical Microcavities},
author = {C. Panuski and D. Englund and R. Hamerly},
journal = {Phys. Rev. X},
volume = {10},
issue = {4},
pages = {041046},
numpages = {28},
year = {2020},
month = {Dec},
publisher = {American Physical Society},
doi = {10.1103/PhysRevX.10.041046},
url = {https://link.aps.org/doi/10.1103/PhysRevX.10.041046}
}

@article{Xiang2023,
author={C. Xiang and W. Jin and O. Terra and B. Dong and H. Wang and L. Wu and J. Guo and T. J. Morin and E. Hughes and J. Peters and Q. Ji and A. Feshali and M. Paniccia and K. J. Vahala and J. E. Bowers},
journal = {Nature},
title = {3{D} integration enables ultralow-noise isolator-free lasers in silicon photonics},
volume = {620},
year = {2023},
pages = {78-85},
publisher = {Nature Research},
doi = {10.1038/s41586-023-06251-w},
url = {https://doi.org/10.1038/s41586-023-06251-w}
}

@article{Jin2022,
author = {N. Jin and C. A. McLemore and D. Mason and J. P. Hendrie and Y. Luo and M. L. Kelleher and P. Kharel and F. Quinlan and S. A. Diddams and P. T. Rakich},
journal = {Optica},
number = {9},
pages = {965-970},
publisher = {Optica Publishing Group},
title = {Micro-fabricated mirrors with finesse exceeding one million},
volume = {9},
month = {Sep},
year = {2022},
url = {https://opg.optica.org/optica/abstract.cfm?URI=optica-9-9-965},
doi = {10.1364/OPTICA.467440},
}

@article{Lee2013,
author={H. Lee and M. Suh and T. Chen and J. Li and S. A. Diddams and K. J. Vahala},
journal = {Nat. Commun.},
title = {Spiral resonators for on-chip laser frequency stabilization},
volume = {4},
pages = {2468},
year = {2013},
publisher = {Nature Research},
doi = {10.1038/ncomms3468},
url = {https://doi.org/10.1038/ncomms3468}
}

@article{Guo2022,
author = {J. Guo and C. A. McLemore and C. Xiang and D. Lee  and L. Wu and W. Jin and M. Kelleher and N. Jin and D. Mason and L. Chang and A. Feshali and M. Paniccia and P. T. Rakich and K. J. Vahala and S. A. Diddams and F. Quinlan and J. E. Bowers },
title = {Chip-based laser with 1-hertz integrated linewidth},
journal = {Science Advances},
volume = {8},
pages = {43},
year = {2022},
doi = {10.1126/sciadv.abp9006},
URL = {https://www.science.org/doi/abs/10.1126/sciadv.abp9006},
}

@article{Liu2022,
author = {K. Liu and N. Chauhan and J. Wang and A. Isichenko and G. M. Brodnik and P. A. Morton and R. O. Behunin and S. B. Papp and D. J. Blumenthal},
journal = {Optica},
number = {7},
pages = {770-775},
publisher = {Optica Publishing Group},
title = {36 {H}z integral linewidth laser based on a photonic integrated 4.0 m coil resonator},
volume = {9},
month = {Jul},
year = {2022},
url = {https://opg.optica.org/optica/abstract.cfm?URI=optica-9-7-770},
doi = {10.1364/OPTICA.451635},
}

@article{Ludlow2007,
author = {A. D. Ludlow and X. Huang and M. Notcutt and T. Zanon-Willette and S. M. Foreman and M. M. Boyd and S. Blatt and J. Ye},
journal = {Opt. Lett.},
number = {6},
pages = {641--643},
publisher = {Optica Publishing Group},
title = {Compact, thermal-noise-limited optical cavity for diode laser stabilization at $1 \times 10^{-15}$},
volume = {32},
month = {Mar},
year = {2007},
url = {https://opg.optica.org/ol/abstract.cfm?URI=ol-32-6-641},
doi = {10.1364/OL.32.000641},
}

@article{Notcutt2005,
author = {M. Notcutt and L. Ma and J. Ye and J. L. Hall},
journal = {Opt. Lett.},
number = {14},
pages = {1815--1817},
publisher = {Optica Publishing Group},
title = {Simple and compact 1-{H}z laser system via an improved mounting configuration of a reference cavity},
volume = {30},
month = {Jul},
year = {2005},
url = {https://opg.optica.org/ol/abstract.cfm?URI=ol-30-14-1815},
doi = {10.1364/OL.30.001815},
}

@article{Haffner2008,
title = {Quantum computing with trapped ions},
journal = {Physics Reports},
volume = {469},
number = {4},
pages = {155-203},
year = {2008},
issn = {0370-1573},
doi = {https://doi.org/10.1016/j.physrep.2008.09.003},
url = {https://www.sciencedirect.com/science/article/pii/S0370157308003463},
author = {H. Häffner and C.F. Roos and R. Blatt}
}

@article{Fortier2011,
  title = {Generation of ultrastable microwaves via optical frequency division},
  author = {Fortier, T. M. and Kirchner, M. S. and Quinlan, F. and Taylor, J. and Bergquist, J. C. and Rosenband, T. and Lemke, N. and Ludlow, A. and Jiang, Y. and Oates, C. W. and Diddams, S. A.},
  journal = {Nat. Photonics},
  volume = {5},
  year = {2011},
  pages = {425-429},
  publisher = {Nature Research},
  doi = {https://doi.org/10.1038/nphoton.2011.121},
}

@article{Li2014,
author = {J. Li  and X. Yi  and H. Lee  and S. A. Diddams  and K. J. Vahala },
title = {Electro-optical frequency division and stable microwave synthesis},
journal = {Science},
volume = {345},
number = {6194},
pages = {309-313},
year = {2014},
doi = {10.1126/science.1252909},
URL = {https://www.science.org/doi/abs/10.1126/science.1252909},
eprint = {https://www.science.org/doi/pdf/10.1126/science.1252909}
}

@article{Kudelin2024,
title = {Photonic chip-based low-noise microwave oscillator},
author = {I. Kudelin and W. Groman and Q. Ji and J. Guo and M. Kelleher and D. Lee and T. Nakamura and C. A. McLemore and P. Shirmohammadi and S. Hanifi and H. Cheng and N. Jin and L. Wu and S. Halladay and Y. Luo and Z. Dai and W. Jin and J. Bai and Y. Liu and W. Zhang and C. Xiang and L. Chang and V. Iltchenko and O. Miller and A. Matsko and S. M. Bowers and P. T. Rakich and J. C. Campbell and J. E. Bowers and K. J. Vahala and F. Quinlan and S. A. Diddams},
journal = {Nature},
volume = {627},
year = {2024},
pages = {534-539},
publisher = {Nature Research},
doi = {https://doi.org/10.1038/s41586-024-07058-z},
}

@article{Sun2024,
title = {Integrated optical frequency division for microwave and mmWave generation},
author = {S. Sun and B. Wang and K. Liu and M. W. Harrington and F. Tabatabaei and R. Liu and J. Wang and S. Hanifi and J. S. Morgan and M. Jahanbozorgi and Z. Yang and S. M. Bowers and P. A. Morton and K. D. Nelson and A. Beling and D. J. Blumenthal and X. Yi},
journal = {Nature},
volume = {627},
year = {2024},
pages = {540-545},
publisher = {Nature Research},
doi = {https://doi.org/10.1038/s41586-024-07057-0},
}

@article{Loh2025,
title = {Magic cancellation point for vibration resilient ultrastable microwave signal synthesis},
author = {W. Loh and D. Gray and R. Maxson and D. Kharas and J. Plant and P. W. Juodawlkis and C. Sorace-Agaskar and S. Yegnanarayanan},
journal = {Nat. Commun.},
volume = {16},
year = {2025},
number = {7997},
publisher = {Nature Research},
doi = {https://doi.org/10.1038/s41467-025-63369-3},
}

@article{Rafac2000,
  title = {Sub-dekahertz Ultraviolet Spectroscopy of ${}^{199}${H}g$^{+}$},
  author = {Rafac, R. J. and Young, B. C. and Beall, J. A. and Itano, W. M. and Wineland, D. J. and Bergquist, J. C.},
  journal = {Phys. Rev. Lett.},
  volume = {85},
  issue = {12},
  pages = {2462--2465},
  numpages = {0},
  year = {2000},
  month = {Sep},
  publisher = {American Physical Society},
  doi = {10.1103/PhysRevLett.85.2462},
  url = {https://link.aps.org/doi/10.1103/PhysRevLett.85.2462}
}

@article{Marra2018,
author = {G. Marra  and C. Clivati  and R. Luckett  and A. Tampellini  and J. Kronjäger  and L. Wright  and A. Mura  and F. Levi  and S. Robinson  and A. Xuereb  and B. Baptie  and D. Calonico},
title = {Ultrastable laser interferometry for earthquake detection with terrestrial and submarine cables},
journal = {Science},
volume = {361},
number = {6401},
pages = {486-490},
year = {2018},
doi = {10.1126/science.aat4458},
URL = {https://www.science.org/doi/abs/10.1126/science.aat4458},
eprint = {https://www.science.org/doi/pdf/10.1126/science.aat4458}}

@article{Liu2024,
author = {Y. Liu and N. Jin and D. Lee and C. McLemore and T. Nakamura and M. Kelleher and H. Cheng and S. Schima and N. Hoghooghi and S. Diddams and P. Rakich and F. Quinlan},
journal = {Optica},
number = {9},
pages = {1205--1211},
publisher = {Optica Publishing Group},
title = {Ultrastable vacuum-gap Fabry--Perot cavities operated in air},
volume = {11},
month = {Sep},
year = {2024},
url = {https://opg.optica.org/optica/abstract.cfm?URI=optica-11-9-1205},
doi = {10.1364/OPTICA.532883},
}

@article{WLoh2025,
title = {Optical atomic clock interrogation using an integrated spiral cavity laser},
author = {W. Loh and D. Reens and D. Kharas and A. Sumant and C. Belanger and R. T. Maxson and A. Medeiros and W. Setzer and D. Gray and K. BeBry and C. D. Bruzewicz and J. Plant and J. Liddell and G. N. West and S. Doshi and M. Roychowdhury and M. E. Kim and D. Braje and P. W. Juodawlkis and J. Chiaverini and R. McConnell},
journal = {Nat. Photon.},
year = {2025},
publisher = {Nature Research},
doi = {https://doi.org/10.1038/s41566-024-01588-8},
}

@article{He2024,
author = {Y. He  and L. Cheng  and H. Wang  and Y. Zhang  and R. Meade  and K. Vahala  and M. Zhang  and J. Li },
title = {Chip-scale high-performance photonic microwave oscillator},
journal = {Science Advances},
volume = {10},
number = {33},
pages = {eado9570},
year = {2024},
doi = {10.1126/sciadv.ado9570},
URL = {https://www.science.org/doi/abs/10.1126/sciadv.ado9570},
eprint = {https://www.science.org/doi/pdf/10.1126/sciadv.ado9570}}

@article{Kefelian2009,
author = {F. K\'{e}f\'{e}lian and H. Jiang and P. Lemonde and G. Santarelli},
journal = {Opt. Lett.},
number = {7},
pages = {914--916},
publisher = {Optica Publishing Group},
title = {Ultralow-frequency-noise stabilization of a laser by locking to an optical fiber-delay line},
volume = {34},
month = {Apr},
year = {2009},
url = {https://opg.optica.org/ol/abstract.cfm?URI=ol-34-7-914},
doi = {10.1364/OL.34.000914},
}

@article{Jeon2025,
author = {I. Jeon and W. Jeong and C. Ahn and J. Kim},
journal = {Opt. Lett.},
number = {4},
pages = {1057--1060},
publisher = {Optica Publishing Group},
title = {$10^{-15}$-level laser stabilization down to fiber thermal noise limit using self-homodyne detection},
volume = {50},
month = {Feb},
year = {2025},
url = {https://opg.optica.org/ol/abstract.cfm?URI=ol-50-4-1057},
doi = {10.1364/OL.541281},
}

@article{Cheng2025,
author = {K. Cheng and N. Wei and Y. Zhang and H. Tao and Y. Hu and J. Chen and N. Dong and J. He and J. Wang},
journal = {Opt. Lett.},
number = {6},
pages = {1783--1786},
publisher = {Optica Publishing Group},
title = {On-chip {M}ach-{Z}ehnder interferometer for 1550 nm laser frequency stabilization},
volume = {50},
month = {Mar},
year = {2025},
url = {https://opg.optica.org/ol/abstract.cfm?URI=ol-50-6-1783},
doi = {10.1364/OL.554895},
}

@article{Idjadi2024,
title = {Modulation-free laser stabilization technique using integrated cavity-coupled {M}ach-{Z}ehnder interferometer},
author = {M. H. Idjadi and K. Kim and N. K. Fontaine},
journal = {Nat. Commun.},
volume = {15},
year = {2024},
number = {1922},
publisher = {Nature Research},
doi = {https://doi.org/10.1038/s41467-024-46319-3},
}

@article{Zhang2014,
author = {W. Zhang and M. J. Martin and C. Benko and J. L. Hall and J. Ye and C. Hagemann and T. Legero and U. Sterr and F. Riehle and G. D. Cole and M. Aspelmeyer},
journal = {Opt. Lett.},
number = {7},
pages = {1980--1983},
publisher = {Optica Publishing Group},
title = {Reduction of residual amplitude modulation to $1 \times 10^{-6}$ for frequency modulation and laser stabilization},
volume = {39},
month = {Apr},
year = {2014},
url = {https://opg.optica.org/ol/abstract.cfm?URI=ol-39-7-1980},
doi = {10.1364/OL.39.001980},
}

\clearpage

\section{Methods}

\subsection{Spiral MZ Interferometer Design and Operation}\label{sec:spiralMZConfig}

The spiral MZ interferometer is designed to operate with one input port and two output ports. The second input port can also be used, but the sinusoidal interference pattern will exhibit a reduced extinction due to the asymmetric splitting ratio of the input coupler. The input coupler is set to the ratio of $70\%/30\%$ to account for the propagation loss of 25 meters of spiral delay at 0.15 dB/m. The total loss through the spiral is thus 3.75 dB, which when applied to the $70\%$ port, achieves power balance with the $30\%$ port. The output coupler is designed with a $50\%/50\%$ splitting ratio, which yields maximal fringe depth when the two arms of the interferometer are combined. To reach 25 meters in length, the waveguide is carefully spiraled outwards from the sinusoidal bend at the center. The spacing between waveguide traces is 17.75 $\mu$m, which corresponds to a total loss of 0.4 dB due to cross-coupling between waveguides. This loss is much lower by design compared to the total waveguide loss, which we found earlier to be 3.75 dB. The input and output waveguides of the interferometer are tapered to 2.25 $\mu$m width over 1 mm length to enable efficient coupling to optical fiber. The fabrication details of the spiral delay line are similar to those reported earlier \cite{WLoh2025}.

The spiral interferometric laser requires only a few milliWatts of optical power to form a low-noise lock. The power delivered to the input of the chip is 5.15 mW, and the fiber-coupled power at the output of the chip is 0.5 mW in each of the interferometer ports. Accounting for the loss of the input and output splitters and also of the spiral delay line, we determine the coupling loss to be 2.4 dB per facet. To operate at the point of maximum common-mode noise rejection for the balanced photodetector, we send a small sinusoidal amplitude modulation to the SOA and measure the size of the induced frequency noise peak. The DC offset of the interferometer lock, which controls the power balance in each of the interferometer arms via the laser frequency, is tuned until this peak is minimized. After the interferometer lock is engaged, we engage the amplitude lock with a servo bandwidth of $\sim100$ kHz.

\subsection{Spiral MZ Interferometer Analysis}\label{sec:MZAnalysis}

We calculate and analyze the operating parameters of the spiral MZ interferometer for its use in laser stabilization. We start at the output coupler of the interferometer where the electric fields of both interferometer arms are combined

\begin{equation}
\centering
E = \frac{E_0}{2}e^{j2\pi \nu t}+\frac{E_0}{2}e^{j2\pi \nu \left(t+\tau\right)}
\end{equation}

\noindent Here, $E$ is the combined electric field in one of the interferometer's output arms, $E_0$ is the input field amplitude, $\nu$ is the laser frequency, and $t$ and $\tau$ respectively denote the time variable and the unbalanced interferometer delay time. Note that we have assumed perfect power balance in the two arms of the interferometer and the interferometer to be lossless. Grouping terms and simplifying, we obtain

\begin{equation}
\centering
E = E_0e^{j2\pi \nu t}e^{j\pi \nu \tau}cos\left(\pi \nu \tau \right)
\end{equation}

\noindent Finally, multiplying by the conjugate to convert to optical powers and simplifying using trigonometric identities, we find the total output power in one of the interferometer's output arms $\left( P \right)$ as

\begin{equation}
\centering
P = \frac{P_0}{2}\left( 1+cos \left( 2 \pi \nu \tau \right) \right)
\end{equation}

\noindent The other interferometer output is $P = P_0\left( 1-cos \left( 2 \pi \nu \tau \right) \right)/2$, and thus the two outputs combined sum to $P_0$, the total power input into the system.

From Eq. (3), it is evident that the power varies between zero and $P_0$ with an average power of $P_0/2$. The free spectral range (FSR) of the interferometer response is determined by the periodicity of the sinusoid, i.e. $FSR=1/\tau$. Furthermore, we can also calculate an effective linewidth of the sinusoid by specifying its full-width at half maximum. We find the linewidth to be $1/\left(2\tau\right)$, which is typically not narrow---except at very long interferometer delay lengths---due to the sinusoid's 50$\%$ duty cycle. The conversion slope is calculated from a derivative of Eq. (3) with respect to frequency. After subtracting (assuming the use of balanced detection) and performing this derivative on the difference, we obtain

\begin{equation}
\centering
\frac{dP}{d\nu} = -2\pi P_0 \tau sin\left(2\pi\nu\tau \right)
\end{equation}

\noindent The slope varies along the interferometer sinusoidal response and is highest when $2\pi\nu\tau=\pi/2$, at the center of the sinusoid in Eq. (3).

\subsection{Spiral MZ Interferometer SNR Limit}\label{sec:MZSNR}

The transfer function of the spiral MZ interferometer forms a sinusoid that converts frequency fluctuations of the interrogating laser to a detectable amplitude fluctuation. The ideal stabilization point is at the center of the sinusoid where the slope is maximum. The slope and the level of noise both determine the performance limit of the lock, which we derive below.

We start by writing the total measured power fluctuations in the system and setting the total to zero, as enforced by the servo.

\begin{equation}
\centering
\Delta \nu \frac{dP}{d\nu}+\Delta P = 0
\end{equation}

\noindent Here $\Delta \nu$ denotes the frequency fluctuations of the seed laser, $\frac{dP}{d\nu}$ is the conversion slope of the interferometer, and $\Delta P$ denotes the combined sum of the system's power fluctuations outside of the intended conversion of frequency to amplitude. Fourier transforming this equation to the frequency domain and multiplying by its conjugate to write the result in the format of power spectral densities, we obtain

\begin{equation}
\centering
S_{\Delta \nu}\left(f\right)=S_{\Delta P}\left(f\right) / \left(\frac{dP}{d\nu}\right)^2
\end{equation}

\noindent $S_{\Delta \nu}\left(f\right)$ represents the frequency noise spectral density of the stabilized interferometric laser, and $S_{\Delta P}\left(f\right)$ is the spectral density of the system's power fluctuations. The two are directly related by the square of the interferometer's conversion slope.

In our system, the majority of the power fluctuations are a result of the amplified seed laser's RIN. These fluctuations are greatly suppressed in our balanced detection scheme, which enables the interferometer to outperform optical cavities that would otherwise have higher Q. Writing Eq. (6) in terms of the laser RIN, we find

\begin{equation}
\centering
S_{\Delta \nu}\left(f\right)=P_D^{2}\;\text{RIN}\!\left(f\right) / \left(\frac{dP}{d\nu}\right)^2
\end{equation}

\noindent where $P_D$ is the total optical power on the detector and the frequency dependence of the RIN is made explicit. From Eq. (7), we determine that the overall performance of the interferometer lock is not dependent on optical power as the conversion slope also scales with power. However, this isn't true in the regimes of operation where the RIN itself is power dependent. As an example calculation, for RIN of -150 dBc/Hz, 1 mW of total optical power at the detector, and an interferometer period of 8 MHz, we calculate the locked frequency noise to be 1.6 mHz$^2$/Hz.

\subsection{Spiral MZ Interferometer Noise Response}\label{sec:MZNoise}

The interferometer structure responds to temperature and other perturbations of refractive index or length in a similar manner to that of conventional optical cavities, though this may not be initially obvious. We derive below the relations that govern how the interferometer frequency responds to refractive index perturbations.

Starting from Eq. (3), we introduce a small perturbation to the refractive index of the interferometer, which affects the unbalanced delay time. Under a perturbation, Eq. (3) takes on the form of

\begin{equation}
\centering
P + \Delta P = \frac{P_0}{2}\left( 1+cos \left( 2 \pi \left(\nu + \Delta\nu\right) \left(\tau + \Delta\tau\right) \right)\right)
\end{equation}

\noindent where $\Delta P$, $\Delta\nu$, and $\Delta\tau$ are perturbations to the interferometer's output power, seed laser frequency, and spiral delay time, respectively. If the seed laser is locked to the interferometer fringe, then $\Delta P$ must be maintained at zero, and the frequency of the laser must shift to compensate for changes in the interferometer fringe. Multiplying out the argument of the cosine in Eq. (8), we find there are four terms $\left(2\pi\nu\tau, \; 2\pi\nu\Delta\tau, \; 2\pi\Delta\nu\tau, \; 2\pi\Delta\nu\Delta\tau\right)$. The first term is the steady state response, while the last term may be ignored for small-signal perturbations. Therefore, the laser lock enforces the condition that $\nu\Delta\tau = -\Delta\nu\tau$, or written another way

\begin{equation}
\centering
\frac{\Delta\nu}{\nu} = -\frac{\Delta\tau}{\tau}
\end{equation}

The delay time is more explicitly written as

\begin{equation}
\centering
\tau = \frac{n_gL}{c}
\end{equation}

\noindent where $n_g$ is the group index of the waveguiding medium, $L$ is the unbalanced delay length, and $c$ is the speed of light. Under a perturbation, we modify Eq. (10) to the following form

\begin{equation}
\centering
\tau + \Delta\tau = \frac{\left(n_g + \Delta n_g\right) \left(L + \Delta L\right)}{c}
\end{equation}

\noindent Once again taking out the steady-state response and eliminating perturbative terms that are too small, we obtain

\begin{equation}
\centering
\Delta\tau = \frac{n_g \Delta L + \Delta n_g L}{c}
\end{equation}

\noindent Finally, putting Eqs. (10) and (12) into Eq. (9), we derive that

\begin{equation}
\centering
\frac{\Delta\nu}{\nu} = -\frac{\Delta n_g}{n_g}-\frac{\Delta L}{L}
\end{equation}

\noindent This is the same equation that governs the response of optical cavities to refractive index and length perturbations. Thus, we conclude that the interferometer behaves similarly to resonant cavities for laser frequency stabilization.

\section{Data availability}

The data sets that support this study are available on reasonable request. Source data are provided with this paper.

\section{Code availability}

The code used for analysis and simulations are available on reasonable request.

\clearpage

\end{document}